\documentclass{llncs}
\usepackage[T1]{fontenc}
\usepackage{graphicx}
\usepackage{amsmath,amssymb} 
\usepackage{color}
\usepackage[width=135mm,left=23mm,paperwidth=176mm,height=210mm,top=18mm,paperheight=246mm]{geometry}
\usepackage{booktabs}

\newcommand{\ra}[1]{\renewcommand{\arraystretch}{#1}}

\begin{document}
\pagestyle{headings}

\mainmatter

\title{Improving COVID-19 CXR Detection with Synthetic Data Augmentation}

\author{Daniel Schaudt\inst{1} \and Christopher Kloth\inst{2} \and Christian Späte\inst{1} \and Andreas Hinteregger\inst{2} \and Meinrad Beer\inst{2} \and Reinhold von Schwerin\inst{1}}
\institute{Technische Hochschule Ulm - Ulm University of Applied Sciences\\ 
\email{daniel.schaudt@thu.de, spaete@mail.hs-ulm.de, reinhold.vonschwerin@thu.de}
\and
Universitätsklinikum Ulm - Ulm University Medical Center\\
\email{christopher.kloth@uniklinik-ulm.de, andreas.hinteregger@uni-ulm.de, meinrad.beer@uniklinik-ulm.de}
}

\maketitle

\begin{abstract}
Since the beginning of the COVID-19 pandemic, researchers have developed deep learning models to classify COVID-19 induced pneumonia. As with many medical imaging tasks, the quality and quantity of the available data is often limited. In this work we train a deep learning model on publicly available COVID-19 image data and evaluate the model on local hospital chest X-ray data. The data has been reviewed and labeled by two radiologists to ensure a high-quality estimation of the generalization capabilities of the model. Furthermore, we are using a Generative Adversarial Network to generate synthetic X-ray images based on this data. Our results show that using those synthetic images for data augmentation can improve the model's performance significantly. This can be a promising approach for many sparse data domains.
\end{abstract}

\section{Introduction}
The ongoing COVID-19 pandemic brings many challenges for societies all around the globe. For the healthcare sector, it is important to screen infected patients in an effective and reliable manner. This is especially true in an emergency setting, where patients already experience advanced symptoms. The prevalent test used for COVID-19 detection is the reverse transcription polymerase chain reaction (RT-PCR) \cite{Vogels2020,Udugama2020,Yang2020}. This method has a high false negative rate and the processing requires dedicated personnel and can take hours to days \cite{Arevalo_Rodriguez2020}.

Since chest X-ray (CXR) images of COVID-19 patients show typical findings including peripheral opacities and ground class patterns in the absence of pleural effusion \cite{Arevalo_Rodriguez2020,Kong2020}, they can be used as a first-line triage tool \cite{Rubin2020}. This could speed up the identification process, as CXR images are easy to obtain and rather inexpensive with a lower radiation dose than computed tomography (CT) images. Using deep learning models for detection of COVID-19 prevalence in CXR images is promising, because it eliminates the need for specialized medical staff in an emergency setting. This can further help to alleviate the challenges to the healthcare systems around the world and has the potential to save lives.

In this retrospective study, we are training a deep convolutional neural network (CNN) on the openly available COVIDx V8b dataset \cite{Wang2020} and evaluate the model on local hospital CXR data. We specifically choose this learning framework to assess the generalization abilities of a CNN in the medical imaging context. Since high quality CXR image data is sparse, we see this as the most common use case for models in production.

Furthermore, we are using a modified version of the StyleGAN architecture \cite{Karras2019} to generate synthetic COVID-19 positive and COVID-19 negative CXR images for data augmentation. This is done to offset some negative side effects encountered by a distributional shift between the training and testing data.

\section{Related Work}
There has been a lot of previous work on applying deep learning to CXR images to detect a COVID-19 pulmonary disease \cite{Wang2020,Khan2020,Ucar2020,Keidar2021,Shamout2021}. However, most of the existing work is using publicly available CXR and COVID-19 image data. Most of those images are collected from heterogeneous sources with varying image and label quality, which raises concerns about the quality and valid evaluation of deep learning models \cite{Tartaglione2020,Oakden-rayner2017}.

Generative Adversarial Networks (GANs) \cite{Goodfellow2014} have been used for many applications in the medical imaging domain \cite{Frid_Adar2018,Yi2019,Kazeminia2020}. Some studies show promising results specifically for the CXR and COVID-19 domain \cite{Karbhari2021,Motamed2021}. In contrast to existing work, we integrate differentiable augmentation \cite{Zhao2020} into our GAN architecture. This enables us to train on a very small dataset and still get meaningful results.

\section{Materials and Methods}
Our goal for this work is to correctly detect a COVID-19 pulmonary disease in chest X-ray images on local university hospital study data. Therefore, we train a deep learning model on publicly available COVID-19 image data and evaluate the model based on our study data. We further enhance the amount of available training data by generating synthetic X-ray images. In this section we explain the origin and distribution of the data, as well as the deep learning model and training process.

\subsection{Data}
\begin{figure}
\centering
\includegraphics[height=6.5cm]{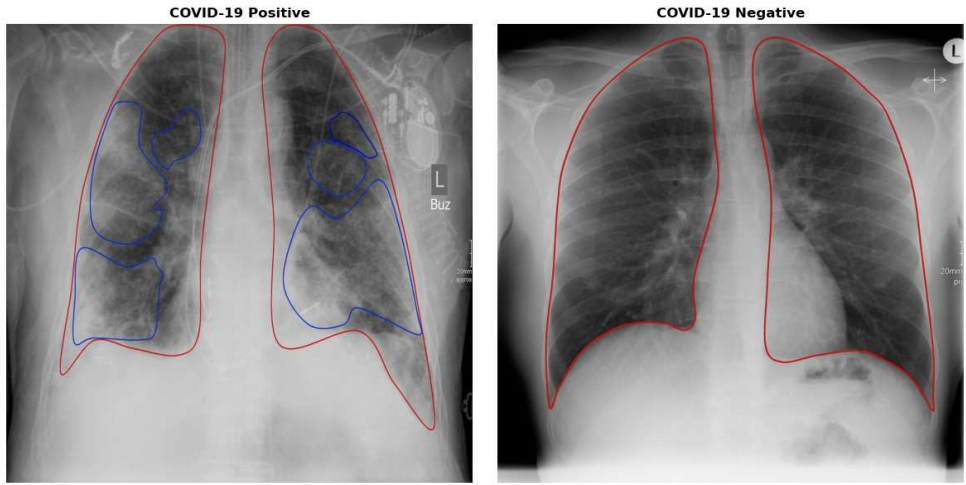}
\caption{Chest X-ray images with lungs marked in red. (Left) COVID-19 positive image, typical Ground-glass opacification marked in blue. (Right) COVID-19 negative image.}
\label{fig:cxr_example}
\end{figure}

In this work we analyze chest X-ray images in posteroanterior (PA) and anteroposterior (AP) front view. Typically the AP view is encountered for cases where the patient is bedridden. Figure \ref{fig:cxr_example} shows two male patient example CXR images from our study data.

\subsubsection{Training data}
\label{subsubsec:training_data}
We use two different training datasets, see Table \ref{tab:data_distribution}. As a first step, we use the COVIDx V8b dataset \cite{Wang2020} to train our model. This dataset is one of the biggest curated and publicly available COVID-19 CXR datasets. We use the training split of the dataset, which contains 13.794 COVID-19 \textit{negative} and 2.158 COVID-19 \textit{positive} frontal view X-ray images of 14.978 unique patients.

In a second step we enhance this training data by using 20.000 additional synthetic CXR images that we generated based on our study data. With that, we can add 10.000 COVID-19 \textit{positive} and 10.000 COVID-19 \textit{negative} images to our existing COVIDx V8b training data. This synthetic data is used to further augment the training of the classification model and increase image diversity. A sample of the generated images has been reviewed by a radiologist to ensure that the model produces meaningful data.

\subsubsection{Validation data}
\label{subsubsec:validation_data}
We validate the model by calculating loss metrics on the so called \emph{test} split of the COVIDx V8B dataset. This dataset contains 200 COVID-19 positive and 200 COVID-19 negative images. We used this dataset to tune model parameters. This is to avoid overfitting our model to the testing data.

\subsubsection{Testing data}
\label{subsubsec:testing_data}
The central data in this work comes from a single center retrospective study of the Universitätsklinikum Ulm. For this study 566 patients (average age 51.12y +/- 18.73y; range 23-82y, 315 women) of a single institution (11/2019-05/2020) were included. The data has been carefully reviewed and labeled by two radiologists after dedicated training into \emph{COVID-19 positive} and \emph{COVID-19 negative}. The senior radiologist (CK) has 8 years of experience in thoracic imaging. This resulted in 110 positive images and 223 negative images, as seen in Table \ref{tab:data_distribution}.

This testing data is used as a holdout set for final model evaluation. With this method we make sure to avoid any patient overlap between the training and testing data. Furthermore, we get a high-quality estimation of the generalization capabilities of the model when evaluating on the testing data. This is because the testing images come from a different data source, which leads to a \emph{distributional shift}.

\begin{table} \centering
\ra{1.3}
\caption{Distribution of images for all datasets}
\label{tab:data_distribution}
\begin{tabular}{lcrr} \toprule
\textbf{Dataset}   & \textbf{Split}  & \textbf{COVID-19 positive} & \textbf{COVID-19 negative} \\ \midrule
COVIDx V8B & Training  & 2.158 & 13.794       \\
COVIDx V8B + Synthetic & Training & 12.158 & 23.794      \\
COVIDx V8B  & Validation & 200 & 200      \\
Uniklinik Ulm Study  & Test & 110 & 223          \\\bottomrule
\end{tabular}
\end{table}

\subsection{Network Architecture}
\label{subsec:network_architecture}
For classification we use the ResNet50 architecture \cite{He2016}. The network has been pretrained on the ImageNet \cite{Deng2009} database. We replace the final fully connected layer with a linear layer of two outputs, one for each class. To get the predictions we apply a softmax activation function. Since training was very stable, we did not use any additional dropout layers or regularization methods.

\begin{figure}
\centering
\includegraphics[height=7cm]{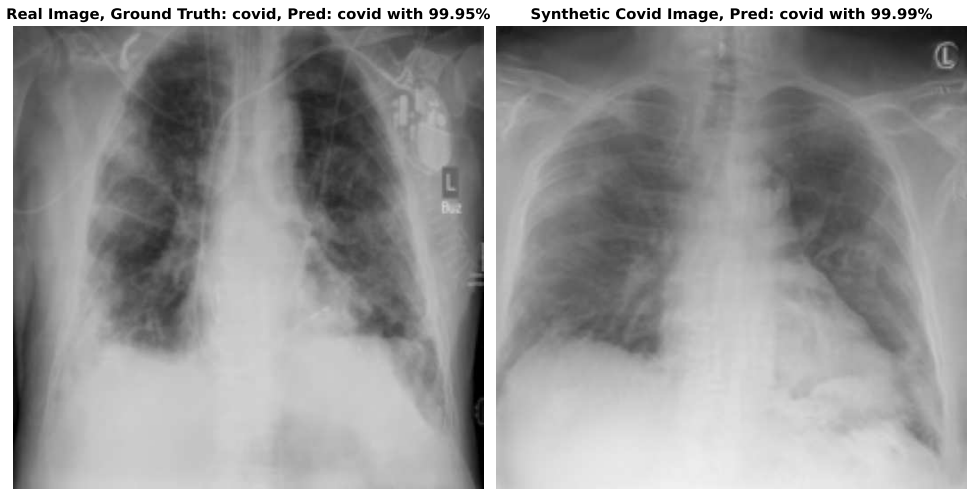}
\caption{Example images for classification by the COVIDx+Synth model. (Left) Real COVID-19 positive image, correctly predicted as \textit{positive}. (Right) Synthetic COVID-19 positive image, correctly predicted as \textit{positive}.}
\label{fig:synth_example}
\end{figure}

The generative model is based on a modified version of StyleGAN \cite{Karras2019}. We specifically integrate differentiable augmentation \cite{Zhao2020} into our StyleGAN architecture. This is to prevent memorization of the training data and helps to stabilize the training process. This combined architecture enables us to generate meaningful synthetic images based on our small study dataset \cite{Spaete2021}. Figure \ref{fig:synth_example} shows one example image along a classification by the COVIDx+Synth model.

\subsection{Training details}
\label{subsec:training_details}
To train the ResNet classifier we use the Adam solver \cite{Kingma2014} with default parameters ($\beta_1 = 0.9$ and  $\beta_2 = 0.999$) and a cross-entropy loss. We train the model using minibatches of size 16. We use an initial learning rate of $0.001$ and apply the One-cycle learning rate scheduler \cite{Smith2017} with a maximum learning rate of $0.006$. We initially freeze all but the new last network layer for 5 epochs of training. After those 5 epochs all network parameters are trained for 30 additional epochs. The One-cycle learning rate scheduler is only applied after the initial freeze period. Increasing the amount of training showed no further improvement empirically.

All images are being scaled down to $224\times224$ and normalized with the mean and standard deviation of images in the COVIDx V8B dataset \cite{Wang2020} before feeding them into the network. During training, we augment the images with random horizontal flipping and random rotation ($\pm5^{\circ}$).

Since we use two different training datasets (see Section \ref{subsubsec:training_data}) we get two different models: \textit{COVIDx} and \textit{COVIDx+Synth}. Both classification models use the exact same hyperparameters and training procedures as described. This is to evaluate the effect of using the synthetic data and make the results comparable.

For the StyleGAN generator, we train two different models: one for each class of COVID-19 positive and negative images. This is a simple method to ensure that we can generate a specific image class. For further details regarding the training process of the StyleGAN generator see Späte 2021 \cite{Spaete2021}.

\section{Results}
To investigate our models in a quantitative manner, we computed the accuracy, as well as F1-score, precision and recall for each class on the validation and testing data. The metrics for the validation data are shown in Table \ref{tab:eval}. Both models perform quite well on the validation data with an accuracy of 96 \% and 95.5 \% respectively. The results are in line with Wang et al. 2020 \cite{Wang2020} and their COVIDNet-CXR-2 model. Interestingly, the Covidx+Synth model falls behind the other models, despite having a lot more training data. This could be another indication of a distributional shift between the COVIDx dataset and the study data of the Universitätsklinikum Ulm.

The results for the testing data are also shown in Table \ref{tab:eval}. The table shows that a model trained on the COVIDx dataset can adapt quite well to the testing data, with an accuracy of 89.49 \%. The model achieves a decent precision for COVID-19 cases (90.32 \%), which is good since too many false positives would increase the burden for the healthcare system due to the need for additional PCR testing. With a rather low recall of 76.36 \% the model does miss quite a lot of COVID-19 cases. This can be especially problematic in this sensitive medical setting, since false negatives lead to undetected cases of COVID-19.

This drawback can be controlled by using additional synthetic data to train the model. Table \ref{tab:eval} shows an increase in accuracy (92.49 \%) and most evaluation metrics. Especially the improved recall of 95.45 \% is very 
desirable. This comes with the cost of a slight reduction in precision (-6.32 \%). Based on those results, it can be seen that our models perform quite well, especially when incorporating the synthetic data, but there are still several areas for improvement.

\begin{table} \centering
\ra{1.3}
\caption{Evaluation metrics for models COVIDx and COVIDx+Synth on validation data (with reported metrics from Wang et al. \cite{Wang2020} for comparison) and on testing data.}
\label{tab:eval}
\begin{tabular}{lccccccc} \toprule
 & \textbf{Accuracy} & \multicolumn{2}{c}{\textbf{F1-Score}} & \multicolumn{2}{c}{\textbf{Precision}}
  & \multicolumn{2}{c}{\textbf{Recall}} \\
\cmidrule(lr){3-4}
\cmidrule(lr){5-6}
\cmidrule(lr){7-8}
\textbf{Model} &  & C19 pos. & C19 neg. & C19 pos. & C19 neg. & C19 pos. & C19 neg. \\ \midrule
\emph{Validation Data} & & & & & & \\
COVIDx & \textbf{0.9600} & \textbf{0.9583} & \textbf{0.9615} & \textbf{1.0000} & 0.9259 & 0.9200 & \textbf{1.0000} \\
COVIDx+Synth & 0.9550 & 0.9548 & 0.9552 & 0.9596 & 0.9505 & 0.9500 & 0.9600 \\ 
COVIDNet-CXR-2 \cite{Wang2020} & - & - & - & 0.9700 & \textbf{0.9560} & \textbf{0.9550} & 0.9700 \\ \midrule
\emph{Testing Data} & & & & & & \\
COVIDx & 0.8949 & 0.8276 & 0.9244 & \textbf{0.9032} & 0.8917 & 0.7636 & \textbf{0.9596} \\
COVIDx+Synth & \textbf{0.9249} & \textbf{0.8936} & \textbf{0.9420} & 0.8400 & \textbf{0.9760} & \textbf{0.9545} & 0.9103 \\ \bottomrule
\end{tabular}
\end{table}

\section{Limitations and Discussion}
In this work we showed that a deep learning model trained with a comparatively large volume of publicly available data for COVID-19 detection is able to generalize well to single source, local hospital data with patient demographics and technical parameters independent of the training data. This is not without limitations, since the distributional shift between the training and testing data can lead to some undesirable results, especially for important metrics like low recall values.

We show that this can be improved by using synthetically generated data to augment the training data. Although this works quite well, one of the reasons could be a rebalancing effect, that could have been achieved with various resampling methods as well. Another reason could be a light form of \emph{data leakage}, since the synthetic data was generated based on the testing data. This is not fully clear, since the StyleGAN generator has no direct access to the ground truth data and just learns based on the feedback of a discriminator. Despite these concerns, the model shows promising first results and using such a model in an emergency setting could give a fast estimation for the prevalence of pulmonary infiltrates and therefore improve clinical decision-making and resource allocation.

\bibliographystyle{splncs}
\bibliography{ms}

\begin{thebibliography}{10}

\bibitem{Vogels2020}
Vogels, C.B., Brito, A.F., Wyllie, A.L., Fauver, J.R., Ott, I.M., Kalinich,
  C.C., Petrone, M.E., Casanovas-Massana, A., Muenker, M.C., Moore, A.J.,
  Klein, J., Lu, P., Lu-Culligan, A., Jiang, X., Kim, D.J., Kudo, E., Mao, T.,
  Moriyama, M., Oh, J.E., Park, A., Silva, J., Song, E., Takahashi, T., Taura,
  M., Tokuyama, M., Venkataraman, A., Weizman, O.E., Wong, P., Yang, Y.,
  Cheemarla, N.R., White, E.B., Lapidus, S., Earnest, R., Geng, B.,
  Vijayakumar, P., Odio, C., Fournier, J., Bermejo, S., Farhadian, S., Cruz,
  C.S.D., Iwasaki, A., Ko, A.I., Landry, M.L., Foxman, E.F., Grubaugh, N.D.:
\newblock Analytical sensitivity and efficiency comparisons of {SARS}-{COV}-2
  {qRT}-{PCR} primer-probe sets.
\newblock (apr 2020)

\bibitem{Udugama2020}
Udugama, B., Kadhiresan, P., Kozlowski, H.N., Malekjahani, A., Osborne, M., Li,
  V.Y.C., Chen, H., Mubareka, S., Gubbay, J.B., Chan, W.C.W.:
\newblock Diagnosing {COVID}-19: The disease and tools for detection.
\newblock {ACS} Nano \textbf{14}(4) (mar 2020)  3822--3835

\bibitem{Yang2020}
Yang, Y., Yang, M., Shen, C., Wang, F., Yuan, J., Li, J., Zhang, M., Wang, Z.,
  Xing, L., Wei, J., Peng, L., Wong, G., Zheng, H., Wu, W., Liao, M., Feng, K.,
  Li, J., Yang, Q., Zhao, J., Zhang, Z., Liu, L., Liu, Y.:
\newblock Evaluating the accuracy of different respiratory specimens in the
  laboratory diagnosis and monitoring the viral shedding of 2019-{nCoV}
  infections.
\newblock (feb 2020)

\bibitem{Arevalo_Rodriguez2020}
Arevalo-Rodriguez, I., Buitrago-Garcia, D., Simancas-Racines, D.,
  Zambrano-Achig, P., Campo, R.D., Ciapponi, A., Sued, O.,
  Mart{\'{\i}}nez-Garc{\'{\i}}a, L., Rutjes, A., Low, N., Bossuyt, P.M.,
  Perez-Molina, J.A., Zamora, J.:
\newblock {FALSE}-{NEGATIVE} {RESULTS} {OF} {INITIAL} {RT}-{PCR} {ASSAYS} {FOR}
  {COVID}-19: A {SYSTEMATIC} {REVIEW}.
\newblock (apr 2020)

\bibitem{Kong2020}
Kong, W., Agarwal, P.P.:
\newblock Chest imaging appearance of {COVID}-19 infection.
\newblock Radiology: Cardiothoracic Imaging \textbf{2}(1) (feb 2020)  e200028

\bibitem{Rubin2020}
Rubin, G.D., Ryerson, C.J., Haramati, L.B., Sverzellati, N., Kanne, J.P.,
  Raoof, S., Schluger, N.W., Volpi, A., Yim, J.J., Martin, I.B.K., Anderson,
  D.J., Kong, C., Altes, T., Bush, A., Desai, S.R., onathan Goldin, Goo, J.M.,
  Humbert, M., Inoue, Y., Kauczor, H.U., Luo, F., Mazzone, P.J., Prokop, M.,
  Remy-Jardin, M., Richeldi, L., Schaefer-Prokop, C.M., Tomiyama, N., Wells,
  A.U., Leung, A.N.:
\newblock The role of chest imaging in patient management during the {COVID}-19
  pandemic: A multinational consensus statement from the fleischner society.
\newblock Radiology \textbf{296}(1) (July 2020)  172--180

\bibitem{Wang2020}
Wang, L., Lin, Z.Q., Wong, A.:
\newblock {COVID}-net: a tailored deep convolutional neural network design for
  detection of {COVID}-19 cases from chest x-ray images.
\newblock Scientific Reports \textbf{10}(1) (November 2020)

\bibitem{Karras2019}
Karras, T., Laine, S., Aila, T.:
\newblock A style-based generator architecture for generative adversarial
  networks.
\newblock In: 2019 {IEEE}/{CVF} Conference on Computer Vision and Pattern
  Recognition ({CVPR}), {IEEE} (June 2019)

\bibitem{Khan2020}
Khan, A.I., Shah, J.L., Bhat, M.M.:
\newblock {CoroNet}: A deep neural network for detection and diagnosis of
  {COVID}-19 from chest x-ray images.
\newblock Computer Methods and Programs in Biomedicine \textbf{196} (November
  2020)  105581

\bibitem{Ucar2020}
Ucar, F., Korkmaz, D.:
\newblock {COVIDiagnosis}-net: Deep bayes-{SqueezeNet} based diagnosis of the
  coronavirus disease 2019 ({COVID}-19) from x-ray images.
\newblock Medical Hypotheses \textbf{140} (July 2020)  109761

\bibitem{Keidar2021}
Keidar, D., Yaron, D., Goldstein, E., Shachar, Y., Blass, A., Charbinsky, L.,
  Aharony, I., Lifshitz, L., Lumelsky, D., Neeman, Z., Mizrachi, M., Hajouj,
  M., Eizenbach, N., Sela, E., Weiss, C.S., Levin, P., Benjaminov, O., Bachar,
  G.N., Tamir, S., Rapson, Y., Suhami, D., Atar, E., Dror, A.A., Bogot, N.R.,
  Grubstein, A., Shabshin, N., Elyada, Y.M., Eldar, Y.C.:
\newblock {COVID}-19 classification of x-ray images using deep neural networks.
\newblock European Radiology (may 2021)

\bibitem{Shamout2021}
Shamout, F.E., Shen, Y., Wu, N., Kaku, A., Park, J., Makino, T.,
  Jastrz{\k{e}}bski, S., Witowski, J., Wang, D., Zhang, B., Dogra, S., Cao, M.,
  Razavian, N., Kudlowitz, D., Azour, L., Moore, W., Lui, Y.W.,
  Aphinyanaphongs, Y., Fernandez-Granda, C., Geras, K.J.:
\newblock An artificial intelligence system for predicting the deterioration of
  {COVID}-19 patients in the emergency department.
\newblock npj Digital Medicine \textbf{4}(1) (may 2021)

\bibitem{Tartaglione2020}
Tartaglione, E., Barbano, C.A., Berzovini, C., Calandri, M., Grangetto, M.:
\newblock Unveiling {COVID}-19 from {CHEST} x-ray with deep learning: A hurdles
  race with small data.
\newblock International Journal of Environmental Research and Public Health
  \textbf{17}(18) (September 2020)  6933

\bibitem{Oakden-rayner2017}
Oakden-Rayner, L.:
\newblock Exploring the chestxray14 dataset: problems (Dec 2017)

\bibitem{Goodfellow2014}
Goodfellow, I., Pouget-Abadie, J., Mirza, M., Xu, B., Warde-Farley, D., Ozair,
  S., Courville, A., Bengio, Y.:
\newblock Generative adversarial nets.
\newblock In Ghahramani, Z., Welling, M., Cortes, C., Lawrence, N., Weinberger,
  K.Q., eds.: Advances in Neural Information Processing Systems. Volume~27.,
  Curran Associates, Inc. (2014)

\bibitem{Frid_Adar2018}
Frid-Adar, M., Diamant, I., Klang, E., Amitai, M., Goldberger, J., Greenspan,
  H.:
\newblock {GAN}-based synthetic medical image augmentation for increased {CNN}
  performance in liver lesion classification.
\newblock Neurocomputing \textbf{321} (dec 2018)  321--331

\bibitem{Yi2019}
Yi, X., Walia, E., Babyn, P.:
\newblock Generative adversarial network in medical imaging: A review.
\newblock Medical Image Analysis \textbf{58} (dec 2019)  101552

\bibitem{Kazeminia2020}
Kazeminia, S., Baur, C., Kuijper, A., van Ginneken, B., Navab, N., Albarqouni,
  S., Mukhopadhyay, A.:
\newblock {GANs} for medical image analysis.
\newblock Artificial Intelligence in Medicine \textbf{109} (sep 2020)  101938

\bibitem{Karbhari2021}
Karbhari, Y., Basu, A., Geem, Z.W., Han, G.T., Sarkar, R.:
\newblock Generation of synthetic chest x-ray images and detection of
  {COVID}-19: A deep learning based approach.
\newblock Diagnostics \textbf{11}(5) (may 2021)  895

\bibitem{Motamed2021}
Motamed, S., Rogalla, P., Khalvati, F.:
\newblock {RANDGAN}: Randomized generative adversarial network for detection of
  {COVID}-19 in chest x-ray.
\newblock Scientific Reports \textbf{11}(1) (apr 2021)

\bibitem{Zhao2020}
Zhao, S., Liu, Z., Lin, J., Zhu, J.Y., Han, S.:
\newblock Differentiable augmentation for data-efficient gan training.
\newblock In: Conference on Neural Information Processing Systems (NeurIPS).
  (2020)

\bibitem{He2016}
He, K., Zhang, X., Ren, S., Sun, J.:
\newblock Deep residual learning for image recognition.
\newblock In: 2016 {IEEE} Conference on Computer Vision and Pattern Recognition
  ({CVPR}), {IEEE} (June 2016)

\bibitem{Deng2009}
Deng, J., Dong, W., Socher, R., Li, L.J., Li, K., Fei-Fei, L.:
\newblock {ImageNet}: A large-scale hierarchical image database.
\newblock In: 2009 {IEEE} Conference on Computer Vision and Pattern
  Recognition, {IEEE} (jun 2009)

\bibitem{Spaete2021}
Spaete, C.:
\newblock Synthetic generation of medical images (unpublished master's thesis).
\newblock Master's thesis, Technische Hochschule Ulm, Ulm (2021)

\bibitem{Kingma2014}
Kingma, D.P., Ba, J.:
\newblock Adam: A method for stochastic optimization.
\newblock (2014)

\bibitem{Smith2017}
Smith, L.N., Topin, N.:
\newblock Super-convergence: Very fast training of neural networks using large
  learning rates.
\newblock (2017)

\end{thebibliography}
\end{document}